# Deep Learning for Medical Image Registration: A Comprehensive Review

**Subrato Bharati [1], M. Rubaiyat Hossain Mondal [2], Prajoy Podder [3], V. B. Surya Prasath [4]**

[1,2,3] Institute of Information and Communication Technology, Bangladesh University of Engineering and Technology, Dhaka-1205, Bangladesh
[4] Division of Biomedical Informatics, Cincinnati Children's Hospital Medical Center, Cincinnati 45229, USA
[4] Department of Pediatrics, University of Cincinnati College of Medicine, Cincinnati, OH 45257 USA
[4] Department of Biomedical Informatics, College of Medicine, University of Cincinnati, Cincinnati, OH 45267 USA
[4] Department of Electrical Engineering and Computer Science, University of Cincinnati, OH 45221 USA
[1] *subratobharati1@gmail.com*, [2] *rubaiyat97@iict.buet.ac.bd*, [3] *prajoypodder@gmail.com*, [4] *prasatsa@uc.edu*

*Abstract*: Image registration is a critical component in the applications of various medical image analyses. In recent years, there has been a tremendous surge in the development of deep learning (DL)-based medical image registration models. This paper provides a comprehensive review of medical image registration. Firstly, a discussion is provided for supervised registration categories, for example, fully supervised, dual supervised, and weakly supervised registration. Next, similarity-based as well as generative adversarial network (GAN)-based registration are presented as part of unsupervised registration. Deep iterative registration is then described with emphasis on deep similarity-based and reinforcement learning-based registration. Moreover, the application areas of medical image registration are reviewed. This review focuses on monomodal and multimodal registration and associated imaging, for instance, X-ray, CT scan, ultrasound, and MRI. The existing challenges are highlighted in this review, where it is shown that a major challenge is the absence of a training dataset with known transformations. Finally, a discussion is provided on the promising future research areas in the field of DL-based medical image registration.

*Keywords*: Deep learning, image registration, medical imaging, unsupervised, supervised, iterative registration, multimodal imaging

## I. Introduction

Using image registration, it is possible to merge disparate picture collections into a single coordinate system with identical information. When comparing two images that were taken from different angles, at several times, or using various modalities/sensors, registration can be required [1, 2]. Until in recent times, the majority of image registration was done by doctors manually. Manual alignments are largely relied on the user's capability, which might be clinically detrimental to the quality of certain registration procedures. Automatic registration was generated to overcome some of the possible disadvantages of manual image registration. The DL renaissance has transformed the background of the research on image registration [3], despite the fact that various approaches to automated image registration have been intensively investigated before (and during). DL [4] has enabled the performance of recent work in a wide range of computer vision tasks, including but not limited to: image classification [4], segmentation [5], feature extraction [6-8], and object recognition [9]. As a starting point, DL proved to be useful in enhancing the performance of intensity-based registration. It was just a matter of time until other researchers looked at the applications of the registration process using reinforcement learning [10-12]. There has been a growing interest in developing unsupervised frameworks for one step transformation estimates due to the difficulty of procuring/creating ground truth data [13, 14]. Image similarity quantification is a well-known roadblock in this paradigm [15]. The application of similarity metrics based on information theory [13], frameworks of generative adversarial network (GAN) [16], and segmentation of anatomical features [17] to solve this difficulty has shown promising results.

Traditional image registration is an iterative-based procedure that involves collecting the necessary features and determining a similarity measure (to assess the registration quality), selecting a model of transformation and lastly a search mechanism [18, 149, 153]. There are two types of pictures that may be sent into the system: moving and fixed, as shown in Fig. 1. It is possible to get the best alignment by sliding the moving picture over the stationary image over and over again. The considered similarity measure initially determines the degree of resemblance between the inputted photos that have been examined. Calculating the new transformation's parameters is done via an optimization method employing an update mechanism. An image with improved alignment is produced by putting these factors into action on the moving image. Otherwise, a new iteration of the algorithm is initiated. If the termination requirements are met, the process is ended. Until no more registration can be obtained or certain predetermined requirements are met, the moving picture improves its correspondence with the stationary image with each cycle. Either the transformation parameters or the final interpolated fused picture may be the system's output.

There is a need for a thorough review of the area of medical image registration using DL, highlighting frequent issues that specialists confront as well as discussing forthcoming research possibilities that can solve these challenges. It is a kind of machine learning (ML) that employs neural networks (NNs) with several layers to learn depictions of data using DL. Many different kinds of neural networks may be utilized for different purposes, and there have been significant designs developed lately to address engineering challenges. There are also many training procedures for neural networks that can be discussed while talking about neural networks. Sections on NN types, training paradigms, network structures, as well as methodologies make up this introduction to DL. PyTorch [19], Caffe [20], Keras [21], MXNet [22], and TensorFlow [23] are all publically accessible libraries that may be applied to create the networks. The existing literature focuses on medical image analysis using DL, reinforcement learning, and GANs in medical image analysis.

This paper provided a comprehensive review of the existing literature on DL-based image registration. The review focuses on the innovations from a methodological and functional perspective. Different forms of registration including unsupervised and supervised transformation estimation, as well as deep iterative registration, are examined in this paper. A discussion is provided on the current trends, challenges, and limitations of image registration. This paper concluded by providing insights into the possible directions for future research.

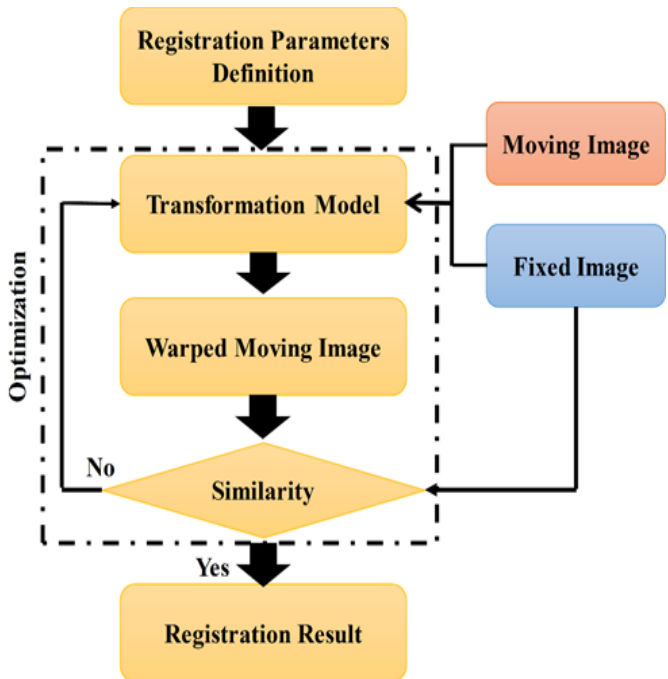

**Figure 1.** An image registration framework flowchart for the medical images

## II. Supervised registration models

For DL models, supervised training is a common foundation for various registration models. There are three sub-categories of models according to the degree of supervision utilized at the stage training: fully supervised, dual-supervised and weakly supervised, and. Fully supervised registration makes use of ground truth DVFs from traditional algorithms of registration to watch over the learning process. These losses are often based on a mismatch between ground truth as well as expected DVFs, as seen in Fig. 2. Rather than using reference DVFs, that are the very widely utilized anatomical contours, the weakly supervised registration uses implicit reference labels, as also seen in Fig. 2. More than two types of reference data are frequently utilized to train dual supervised registration models. This includes anatomical structure contours, reference DVFs as well as image similarity. Below is a summary of the many supervised registration models.

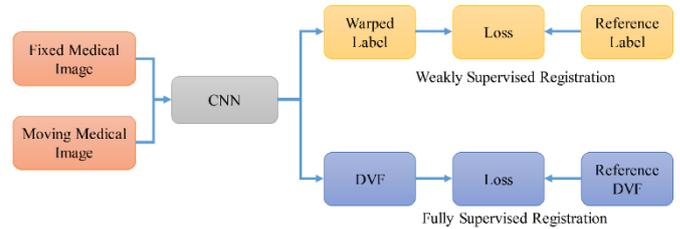

**Figure 2.** A sample working diagram of weakly and fully supervised registration models

### A. Applications of Fully Supervised Registration

DVF denseness was derived via interpolation in a few experiments that predicted movements at patch centers. The output layer of a CNN based model with its 3 neurons was used by Cao et al., where each neuron represented the amplitude of motion along the $x$, $y$, and $z$ axes in the middle of a tiny patch [24]. The authors were able to get results that were as accurate as more traditional methods. Inter-phase registration of lung 4D-CBCT and 4D-CT images was accomplished by the authors of [25] applying a patch-based NN. The deformable vectors in the moving patch centers were predicted using moving and target patch pair inputs. The diaphragm region's registration performance was assessed using three assessment metrics: CC, MSE, and SSIM. In other experiments, CNNs were used to forecast dense DVFs in advance. For instance, Yang et al. learned DVFs with brain MR images of the same resolution using a CNN based U-Net model and obtained excellent registration accuracy across various datasets [26]. A similar network design was used by Rohé et al. [27] for the registration of cardiac MR images, and the results were equivalent to traditional approaches in terms of contour overlap [27]. For smoothing diffeomorphic changes in brain MR images, Wang et al. [28] developed a method for tuning the regularization parameter [28]. The regularization parameters were discovered using pairwise image registration using a CNN prediction model. An effective and memory-saving method of predicting regularization parameters was used by the network's developers. Some works on supervised image registration employed fake DVFs to oversee the training of DL models in order to overcome the lack of training data. The collection of highly labeled data and the supervision of training at the voxel level would both be much easier and more cost effective if DVFs were artificially manufactured. For registering lung CT images, the authors of [29, 30] employed a fully supervised DIR technique, that the DVFs were previously produced to simulate both big and tiny movements. This approach was able to successfully register several datasets. Artificial DVFs may also be created in a variety of ways. These approaches all had registration accuracy that was on par with or better than traditional algorithms, as measured by TRE and the Dice score [31]. Reference deformations may not only improve accuracy, but they can also give biomechanical data and improve the viability of dynamical systems models. Prostate image registration frameworks have been devised that ROIs are 1st segmented, and after that volumetric point in clouds are created from this meshing, utilizing tetrahedron meshing. Using biomechanical constraints, finite-element modeling was used to establish reference deformations from these points. When tested with DSC, MSD, HD, and TRE, the registration framework showed promising registration performance. Fu et al. employed this system to register multi-parametric MR images with CBCT in order to show its generalizability [32]. This strategy outperformed the typical intensity-based rigid registration when it came to TRE scores.

*B. Applications of Dual Supervised Registration*

The use of reference DVFs may speed up the monitoring process, but they are not infallible in all situations. Models trained just with reference DVFs will never outperform those trained using DLs. Using a fully convolutional network with dual guidance, Fan et al. [33] have registered brain MR images in order to correct for incorrect DVFs in previous work [33]. Euclidean distance from predicted to reference DVFs as well as MSE in fixed/warped images were used to evaluate network performance. Hierarchical loss and gap filling features were added to their model in order to boost performance. In addition, they used a variety of data sources to supplement their training materials. On a wide range of datasets, their technique demonstrated promising registration accuracy and efficiency compared to current best practices. The registration of brain MR images was shown by Ahmad et al. in two steps [34]. In their technique, the input photos were exemplified as a graph and grouped using iterative graph roughening prior to performing DL training. As a result, large-scale picture analyses could be conducted more quickly and accurately using this deformation initialization, which was much quicker and more accurate than previous approaches. For large-scale abdominal CT picture distortion, Ha et al. [35] used supervised learning to construct a notion [35]. Distinct heatmaps were predicted using deformable-field and graph convolution for relative displacement between two scans. The sparse displacement between two scans was estimated using smoothness of transformations as well as the MSE of DVF. Compared to state-of-the-art DL techniques for abdominal CT, our method exhibited a significant increase in accuracy.

Several different ROIs have been successfully registered using the fully supervised registration approach. Certain difficult registrations, such as multi-modality as well as big motion registration, have been made possible with performance efficiencies equivalent to traditional approaches thanks to conventional registration methods' reference deformations.

Biomechanical constraints may be learnt using specially prepared training samples, such as reference DVFs, in fully supervised registration. Full supervised registration has several benefits, but the lack of training data significantly restricts its use. Using artificial DVFs and data augmentation methodologies, this issue may be solved. Moving and stationary pictures may be tracked precisely using these two approaches, eliminating the uncertainty generated by reference deformations. It is especially relevant to multi-modality registrations, when reference deformations are less reliable. However, both of these methods may not accurately depict the actual movements of the human body. There is, therefore, a need for coordinated efforts to develop more realistic training samples. Another crucial subgroup is that of registration, with little supervision. The organ contours, in contrast to reference DVFs, are easier to produce and less prone to error. This makes it easy to undertake unsupervised instruction. Contours might potentially be used to handle challenging registration issues, such as loud labels and large motions, because of the supervision they give [36, 37]. It is envisaged that dual supervision will outperform both of the above-described ways of supervision. Dual-supervision sample preparation is more time-consuming. When non-differentiable biomechanical constraints are included, supervised registration has been shown to be a particularly effective method. We anticipate supervised registration to evolve further in the future, given its potential.

*C. Applications of Weakly Supervised Registration*

Several researchers in the field of brain MR-MR registration have embraced the concept of segmentation as part of registration training. Inter- and intra-subject brain MR image registration was achieved using a deformable registration strategy that incorporated global and local labeled learning with CNN [38]. For example, the authors of [39, 40] constructed CNNs based on hybridization methods that were capable of both image segmentation and registration in the same framework. As a result of the commonality in segmentation, registration was made easier and more accurate. Another method, used by Xu et al. [40], was to include the current segmentation into the network as an input [40]. Unsupervised registration was the major emphasis of Balakrishnan et al. [41] although they also included a poor supervision option using contours [41]. All of the above-mentioned networks were able to successfully register data from a wide range of datasets. The authors of [42] also employed a registration system that combines deformable and affine MR to MR brain registration approaches in addition to pure deformable registration. The loss function for the affine network was global similarity, while the loss function for the deformable network was local. Furthermore, the registration network's training was supervised using anatomical similarity as a whole. When compared to other approaches, this one excelled them all.

CT image registration using weakly supervised registration has also proved effective. Using noisy segmentation and spatial gradients labels, the authors of [36] applied a registration approach to CT-CT registration for abdominal. So, for both source-to-target and target-to-source transformations, they devised a formula. They also included public datasets for their training. To accomplish large-scale registration of CT-CT lung, the authors of [37] developed a multilayer variational image registration network. In comparison to traditional procedures, their multi-level strategy was capable of providing much superior outcomes of registration.

## III. Unsupervised Registration Models

The creation of training samples is still a time-consuming procedure, despite the adoption of numerous strategies (such as weak supervision and data augmentation) [43-47] to solve the information or data scarcity issue of supervised image registration. Since moving and fixed picture pairings are all the DL model needs to learn about deformation, unsupervised registration is the way to go. Table 1 provides an overview of this subcategory. A loss function that is comparable to that used in traditional iterative registration is still required for training in this category. A DVF regularization term and an image similarity term and are often included in the loss function. Some similarity measures, i.e., localized NCC (LNCC), are changed to concentrate on tiny patches because of the nature of intrinsic convolution. Various loss terms may be introduced, i.e., identity loss to prevent overfitting and cycle-consistency loss to minimize singularity. As illustrated in Fig. 3(a), a discriminator rather than intensity-based metrics is applied to measure the similarity between warped and fixed images. For deep similarity-based registration, voxel intensities alone are not enough; underlying textures and information must also be taken into account. Fig. 3(b) depicts the overall procedure of unsupervised similarity-based registration. Unsupervised medical image registration using GANs is a subset of this technique.

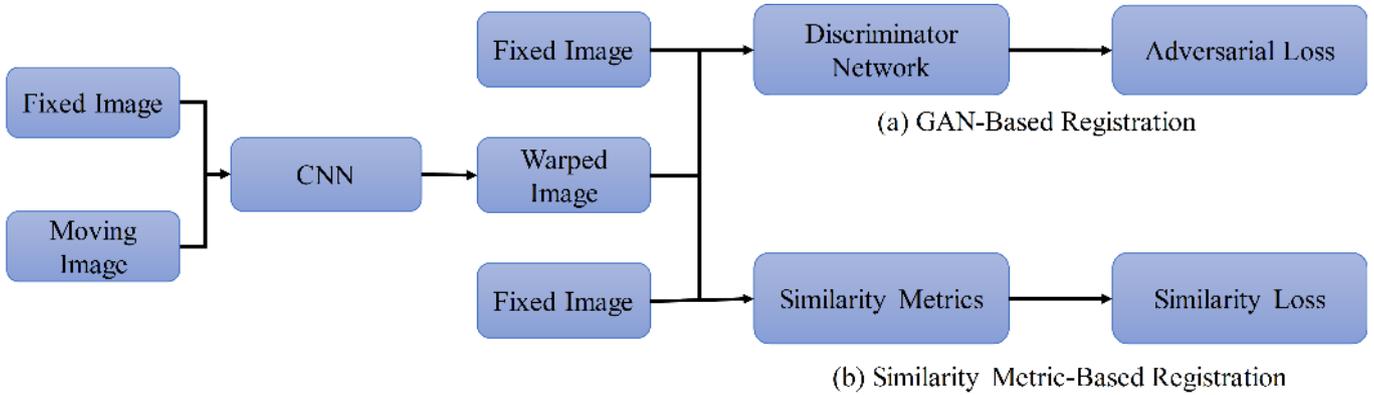

**Figure 3.** Generic framework of medical image registration based on (a) GAN and (b) similarity matrix

*A. Similarity metric-based registration*

DIR of brain MR images was performed by the authors of [41] using an encoder-decoder-like network called VoxelMorph. Inputs for the network were paired brain MR images, which predicted dense DVFs. Based on the installation of the imaging modalities, MSE or NCC were used to monitor network training. The spatial gradient of the DVFs was used to improve the DVFs' believability. A similar decoder-encoder-like network, called U-ReSNet, was constructed by Estienne et al. [48] in order to register brain MR images. The segmentations were performed by extracting medical images characteristics applying a common encoder in a model that was similarly utilized for reconstructing and registering anatomical labels deploying a single decoder in their network, which was innovative. A more accurate registration was made possible thanks to the segmentation findings. Some studies changed the training technique to a cycle-consistent one in order to improve DVF regularity. To do this, they used a network to analyze the distorted picture and convert it back to a moving medical image [49, 53, 58]. As a result, the Jaco. Det. included fewer false negatives, allowing for the generation of more believable DVFs. An identity loss component was included by Kim et al., which punished any distortion of identical images [49]. Researchers often use numerous registration stages and affine registration to accommodate big movements. De Vos et al. [13] used a multi-resolution and multi-level approach to accomplish affine and deformable registration [13]. Multiple steps of downsampling of the original source photos allowed them to catch both huge and tiny movements. Other research has used a comparable coarse-to-fine method for lung imaging as well as obtained good registration accuracies [52, 54, 59]. Their DL models were able to handle substantial deformations in prostate and knee pictures by using affine registration [55, 60]. It is possible to assess their similarities as well as extract features using pre-trained CNNs instead of basic intensity metrics. Perceptual loss is a method that some researchers have utilized in distinct networks to learn deep metrics for improved registration. It has been shown, for example, that the deep similarity between the CT and CBCT images may be learned using a network of spatial weighting-based metrics, as has been established by Duan et al. [61].

*B. GAN-based registration*

For many years, the task of registering MRUS images has been viewed as a problem due to the considerable range in image correspondence and the substantial disparities in the appearance of the images. It was possible to discriminate between warped and fixed pictures using a discriminator and a generator. When compared to traditional approaches, they found that this strategy produced much higher DSCs and significantly lower TREs following registration. Using a shallow discriminator, Elmahdy et al. [62] proved the possibility of performing combined prostate CT DIR and segmentation using a shallow discriminator [62]. To record movements at many scales, former researchers cropped the medical images to use dilated convolutional layers and to get tiny patches to generate and discriminate [52, 59]. An improved discriminator was created by Fan et al. [63] that was capable of receiving two pictures rather than a single one [16, 50]. This was done to alleviate the impracticality of perfect matching, which they described as a mix of a fixed and moving pair of images. Because deformations do not occur in a uniform manner across the body, concentrating on regions that are more prone to bigger deformations may help increase registration accuracy. DL models by the works of [59] contained attention modules to give areas with big movements larger weights. The researchers then cropped photos into tiny patches and categorized them into "difficult" and "easy" patches depending on the amount of attention each patch drew [64]. After categorization, the difficult-to-register patches were fine-tuned.

A major advantage of unsupervised registration is the ease with which it may be taught compared to supervised registration. This has led to an increase in the number of articles on unsupervised registration [65, 66]. Researchers from many different organizations have attained accuracy levels equivalent to or even exceeding those of more traditional methods [52, 54]. Unsupervised multi-modal registration, on the other hand, has received less attention in the past, and hence merits more study. Because anatomical contours or no reference DVFs are supplied, unsupervised medical image registration is inherently more difficult than supervised medical image registration. Accordingly, effective registration requires additional steps beyond basic network training. For example, rigid registration was used by the authors [58] before the DL algorithm was used to minimize motion amplitudes in medical images, while binary masks were used by other researchers to concentrate on ROIs [54]. Using segmentation and a 1,000-fold increase in pulmonary vessel intensity, the author of [52] improved the image's fine features. Preprocessing may help with registration accuracy, but it can also complicate training and limit the generalizability of the model. Finally, unsupervised registration approaches are simple to learn and show great promise in terms of accuracy. This sub-category, therefore, is expected to see an increase in study interest.

| Modality | Similarity Loss | GAN-based | Transform | ROI | Evaluation metrics | Reference |
|---|---|---|---|---|---|---|
| CT-MRI | NCC | No | Deformable | Prostate | ASSD, DSC | [24] |
| MRI-MRI | NCC/MSE | No | Deformable and Affine | Brain | DSC | [48] |
| CT-CT | CC | No | Deformable | Liver | Jaco. Det., TRE | [49] |
| MR-US | N/A | Yes | Affine | Prostate | TRE, DSC | [50] |
| PET-PET | CC | No | Deformable | Chest | MSE | [51] |
| CT-CT | NCC | Yes | Deformable | Lung | TRE | [52] |
| MRI-MRI | CC | No | Deformable | Brain | Jaco. Det., DSC | [53] |
| CT-CT | NCC | No | Deformable | Lung | TRE | [54] |
| MRI-MRI | NCC | No | Deformable and Affine | Knee | DSC | [55] |
| MRI-MRI | NCC | No | Deformable | Brain | DSC | [14] |
| PET-CT | NCC | No | Deformable | Body | NCC | [56] |
| CT-CT | CC | No | Deformable and Affine | Liver | Jaco. Det., DSC | [57] |

*Table 1.* Summary of unsupervised registration

*NCC: normalized cross-correlation; CC: cross-correlation; MSE: mean square error; ASSD: average symmetric surface distance; DSC: Dice coefficient; Jaco. Det.: Jacobian determinant; TRE: target registration error

## IV. Deep Iterative Registration

A metric for measuring the similarity between a moving and stationary image, as well as an optimization technique for updating the transformation parameters to optimize the similarity between the images, are required for automatic intensity-based medical image registration. The sum of mutual information (MI), cross-correlation (CC), squared differences (SSD) [67, 68], normalized mutual information (NMI), and normalized cross-correlation (NCC) [67, 68] were often utilized for similar applications prior to the DL renaissance. Intense-based image registration is a direct extension of DL in medical picture registration [69-71]. Since practitioners do registration in an iterative fashion, some researchers subsequently employed a reinforcement learning paradigm to make iterative estimates of a change [10-12, 72]. A breakdown of the two techniques is shown in Table 2. Later, we will look at some more recently discovered approaches that employ deep reinforcement learning-based medical image registration rather than the older methods that rely on deep similarity registration.

| Model | Modality | Transform | ROI | Learning | Reference |
|---|---|---|---|---|---|
| FCN | CT | Deformable | Lung | Metric | [73] |

| Model | Modality | Transform | ROI | Learning | Reference |
|---|---|---|---|---|---|
| 5-layer DNN | CT/MR | Deformable | Head | Metric | [74] |
| 9-layer CNN | CT | Deformable | Thorax | Metric | [75] |
| 5-layer CNN | MR | Deformable | Brain | Metric | [69] |
| 8-layer CNN | MR | Deformable | Prostate | RL agent | [72] |
| LSTM/STN | Metric | Rigid | brain | MR/US Fetal | [76] |
| 5-layer CNN | Metric | Rigid | Abdominal | MR/US | [77] |
| 8-layer CNN | RL agent | Rigid | Spine/cardiac | CT/CBCT | [10] |
| Dueling network | RL agent | Rigid | Spine | MR/CT | [11] |

*Table 2.* Summary of deep iterative registration techniques for medical imaging

### A. Deep similarity-based medical registration

In this section, approaches for developing a similarity measure using DL are discussed. An intensity-based medical image registration system with an optimization algorithm, interpolation technique, and transformation model is used to include this similarity measure. Fig. 4 depicts the general structure of our study. For both training and testing, the solid lines indicate data flows, whereas for training alone, the dashed lines represent data flows. All of the other figures in this article are consistent with this.

For unimodal registration, manually generated similarity measures work pretty well; DL has been applied to develop better metrics. Before moving on to multimodal registration, this section will focus on ways that employ DL to improve the performance of registration processes using unimodal intensity.

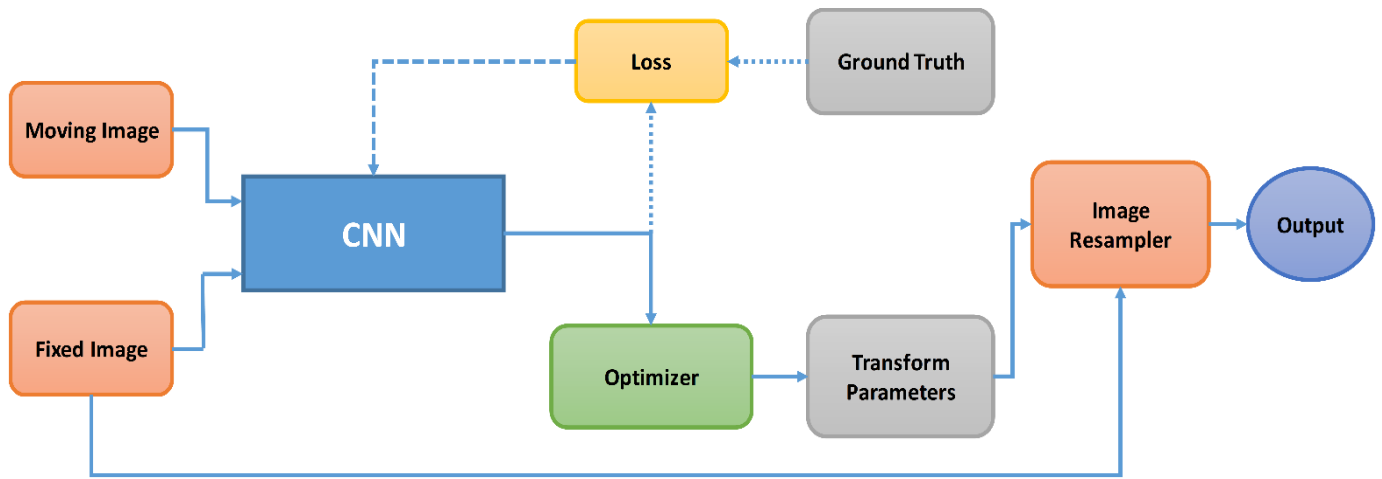

**Figure 4.** A visual representation of the registration process used in research that measure image similarity using DL based framework for an intensity-based registration [133]

*1) Unimodal medical image registration*

First, some authors used DL to develop an application-specific similarity score for medical image registration. For 3D brain MR volume registration, those authors employed a convolutional-stacked autoencoder to generate features for unimodal, deformable registration (CAE). They then used gradient descent to maximize between the features set's two NCCs during the registration process. With this strategy, you may get better results than with diffeomorphic demons and HAMMER [78]. 3D thoracic CT images with its end-to-end deformable registration (inhale to exhale) has recently been approximated by Eppenhof and colleagues [75]. Inhale to exhale pairs of thoracic CT images were applied to generate a 3D CNN error map. This study also relied only on previously learnt characteristics. Lung CT registration might instead make the application of handcrafted MRF-based self-similarity descriptors, and CNN-based descriptors as described by Blendowski et al. [73]. The CNN-based descriptors beat the hand-generated descriptors, but the best results were produced when both sets of descriptors were used. There is evidence that DL may not beat human approaches in the case of unimodal registrations. However, it may be utilized to gather more data.

*2) Multimodal medical image registration*

It is clear that DL can make a big difference in the medical images with its multimodal scenario, while manually built similarity measures have had limited accomplishment. Stacked denoising autoencoders were utilized by Cheng and colleagues [74] to develop an algorithm for measuring the quality of CT-to-MR picture alignment. Local cross correlation (LCC) optimization and NMI optimization were shown to be inferior to their metric in their application. The authors of [69] employed a CNN to learn the dissimilarity from aligned 3D T1 to T2 weighted images in brain MR volumes to explicitly quantify picture similarity in the multimodal situation. Gradient descent was utilized to repeatedly update the parameters of a deformation field based on this similarity measure. This technique outperformed MI-optimization-based registration, paving the way for multimodal registration based on deep intensities in the future. In addition, the authors of. [69] used a 5-layer NN to develop a similarity measure, which was subsequently improved using Powell's approach, to conduct rigid registration images of 3D MR/US. MI-optimization-based registration was also beaten by this method in terms of performance. MRI and Transrectal ultrasound (TRUS) were registered using a similarity metric developed by the authors of [79]. Because of the absence of convexity of the learned metric, they utilized an evolutionary algorithm to discover the solution space before applying a regular optimization approach. In comparison to MI-optimization based and MIND-optimization based [15] registration, this system outperformed.

*B. Discussion and assessment*

New research has recently verified neural networks' capacity to detect picture similarity in multimodal medical image registration. Using the methodologies presented in this section, DL may be effectively used to difficult registration problems. It has been suggested that in the case of unimodal comparisons, learned medical image similarity measures can be the best option. Furthermore, real-time registration is challenging with these iterative approaches.

*1) Applications of Deep similarity-based registration*

SAEs created by Wu et al. [80] have been used to learn and measure the similarity of discriminative characteristics in input images [71]. Since SAEs can learn inherent image properties, the incorporation of SAEs into traditional algorithms has resulted in consistently better registration accuracy across a variety of datasets. Using a neural network, Simonovsky et al. [69] created a classification job to determine if the input picture pairings were properly aligned [69]. They then used CNN to replace the MI in a traditional registration method and found that the resultant algorithm provided considerably better registration of T1–T2 brain MR images. A concept comparable to Sedghi et al. [80] was used, and this groupwise registration approach was further refined by them. The typical MI measure had previously failed to handle challenging registration circumstances, but their deep metric did just as well in such situations. For the registration evaluation, TREs were calculated from the MR-TRUS image pairings. The training was monitored by comparing the actual TREs to the predicted TREs. Conventional MI-based feature-based registration approaches were surpassed by the CNN, which had lower TREs than the current state-of-the-art feature-based approaches. They employed the dissimilarities in image attributes from moving to distorted pictures as the cost function in their pre-trained segmentation network [81]. Compared to traditional approaches, they reported quicker convergence times and better registration accuracy. For rigid and DIR, et al. used Bhattacharyya Distances, a learning-based measure that outperformed MI [82]. The regularization component may also be improved by training, in addition to the similarity term. According to Niethammer et al. [83], a shallow CNN may be used to provide spatial adaptive regularization, for example [83]. For smoothness, shift-invariant deformation field gradient is penalized in conventional transform regularization, which may over smooth areas with abrupt changes.

# V. Reinforcement Learning-based Medical Image Registration

Reinforcement learning (RL)-based medical image registration approaches are examined in this section. In this case, a trained agent rather than a predefined optimization technique is employed to conduct the registration. In Fig. 5, a visual representation of this framework is shown. Most registration models based on reinforcement learning include stiff transformations. However, a deformable transformation model may be used. To date, the authors of [10] are the first to employ RL-based medical image registration to rigidly register abdominal and cardiac 3D CT scans as well as cone-beam CT images. End-to-end training was carried out using an attention-driven hierarchical method with a greedy supervised approach. Registration based on MI as well as semantic medical image registration applying probability maps were both surpassed by this approach. Reinforcement learning was utilized by Kai et al. [11] to register the MR/CT chest volumes shortly after [67]. Based on Q-learning, this technique uses contextual information to estimate the projected image's depth. The dueling network architecture [84] is utilized to construct the network in this manner. Notably, this study makes a distinction between incentives with a finite life span and those that do not. Compared to other approaches, this one surpasses the Dueling Network, Deep Q Network, and iterative closest points (ICP) in registration [84]. Reinforcement learning was utilized to train a multi-agent system to strictly record the spine's CT and X-ray images instead of training a single agent such as the prior approaches [12]. They demonstrated the effectiveness of a multiagent system by observing several locations using an auto-attention method. A recent similarity measure presented by [85] was able to considerably outperform registration systems that utilized it. Deformable registration of prostate MR volumes was performed by the authors [72] using a reinforcement learning-based technique instead of the previous rigid registration-based efforts. Stochastic action selection was controlled by a fuzzy controller and a low-resolution deformation model. In order to keep the action space as two-dimensional as possible, the low-resolution deformation model is required. The LCC Demons [86] and Elastix toolbox [87]-based registration approaches were outperformed by our method. Reward learning is an obvious choice for medical image registration. For reinforcement learning-based registration, handling high-resolution deformation fields is a major difficulty. Rigid registration does not remove any of these difficulties. With their simple and recent origins, we predict that these methodologies will receive greater attention from the scientific community in the coming years.

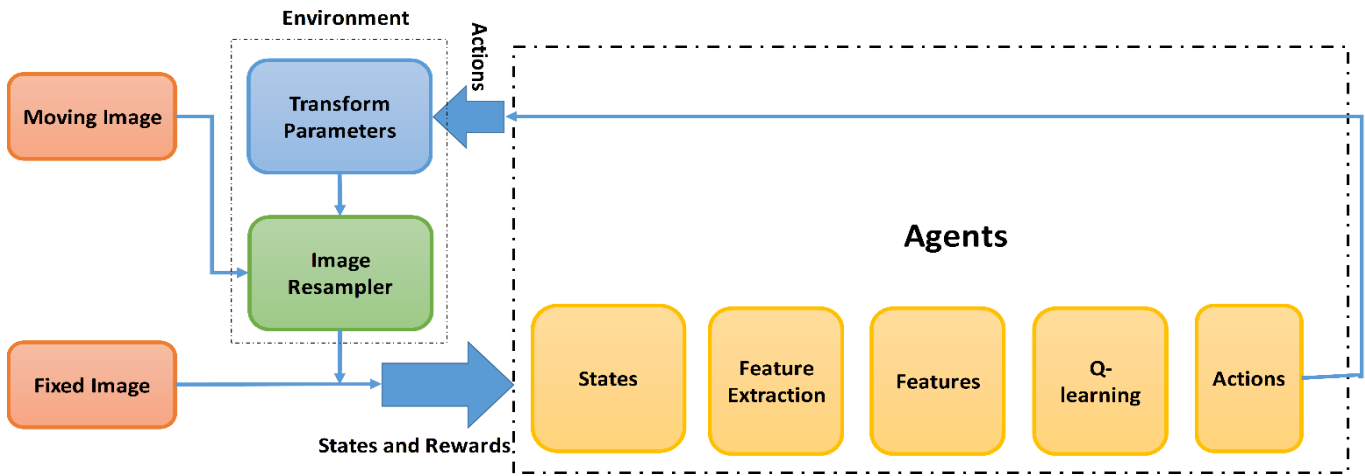

**Figure 5.** A pipeline visualization for studies that employ deep RL to indirectly estimate medical image similarity for medical image registration [72]

For the first time, 3D rigid-body image registration was performed using an RL framework, according to Liao et al. [10]. As part of this method, image alignment is performed using a neural network-based agent that predicts successive motions (e.g., translations and rotations within one millimeter or one rotational angle). Because of the DSL training method, it is not possible to use the AI agent's exploration history to boost training efficiency. A model like statistical deformation was used by the authors of [72] to limit the action space's dimensions, as opposed to the rigid registration method previously discussed. Asynchronous RL was recently used for 2D affine registration by Hu et al. [88]. Convolutional long-short-term memory (conLSTM) was used to extract spatiotemporal picture characteristics from the RL framework.

## VI. Image Based Applications of Registration

Here, we look into DLIR approaches from a new angle, focusing on how they might be put to use. There are several clinical applications that need medical image registration, including illness treatment and diagnosis planning, surgical procedures and image-guided therapy, patient prognostication, and treatment assessment, among many more uses of medical image registration. Real-time compensation for patient motion and soft tissue distinguishes DLIR systems from typical iterative registration procedures. Real-time cardiac motion analysis using this method might lead to the identification of new disease biomarkers. Atlases of population-averaged medical pictures may be estimated using DLIR algorithms. An atlas that is conditional on a variety of variables, including age and gender, was developed by the authors of [89] using a model like probabilistic spatial deformation that built on diffeomorphism. Using appropriate variables, they may also be used to examine the anatomical variability of populations. It is also possible to employ picture registration to directly aid with image segmentation. Bayesian segmentation was developed by Dalca and colleagues [90] using an unsupervised DLIR framework to convert 3D brain MRI [91] from an annotated atlas, eliminating the requirement for laborious manual segmentation of several images. As a result of these investigations, we can see how useful DLIR approaches may be in a variety of contexts.

### A. Monomodal Registration

Table 3 summarizes all publicly accessible datasets used to build the DLIR registration technique, with hyperlinks to each. This will aid future studies on DLIR. There has been a considerable rise in the amount of research focusing on monomodal registrations in the last year. According to the observed trend, the development of DL-based multimodal registration approaches is expected to increase significantly in the next few years. MRI, X-ray, US, and CT are the most prevalent imaging modalities in the clinical setting, hence in this part we concentrate on monomodal DLIR approaches.

| Organ | Modality | Registration | Datasets | References |
|-------|----------|--------------|----------|------------|
| Heart | Cine MRI | Monomodal | Sunnybrook [92] | [13, 93-95] |

| Organ | Modality | Registration | Datasets | References |
|---|---|---|---|---|
| Liver | CT, X-ray | Monomodal | COPDGen [96], LiTS, SLIVER, MSD | [57, 97] |
| Brain | MRI | Monomodal | HABS [98], MCIC [99], ADNI [100], PPMI [101], ABIDE [102] | [41, 89, 90, 103-105] |
| Knee | MRI, X-ray | Multimodal | OAI | [40, 55, 106] |

*Table 3.* Datasets of medical imaging for image registration

*1) CT registration*

Organs in the abdomen and chest (such as the lungs and heart) may be scanned using CT imaging (liver, kidneys, and pancreas). There are 4 liver CT image datasets (LiTS, MSD, SLIVER) and 8 thoracic CT image datasets (Table 3) to choose from DIR-Lab-4DCT [107], DIR-Lab-COPDgen [108], NLST [109], COPDGen [110], Empire 10 lung datasets, POPI [111], LIDC-IDRI [112]. There are also a number of multi-modal medical image datasets that include CT scans, VISCERAL Anatomy 3, RIRE and MM-WHS. Many recent studies have shown that CT scan registration is the second most popular sector for emerging medical image registration algorithms [13, 95, 113-115]. CT image registration is more difficult than brain MRI registration because of the higher variability and the lack of soft-tissue contrast in picture quality.

*2) X-ray registration and Ultrasound (US) registration*

There are not many publicly accessible datasets for X-ray and US images [7, 8, 116-123, 154], contrary to the other imaging modalities we have explored up to now. As a result, there are not many articles out there dealing with the registration of US and X-ray images. Only one work [124] focuses on monomodal US registration utilizing publicly accessible datasets from the 2 brain datasets, BITE and RESECT, which comprise ultrasound images. There are six publicly accessible datasets for X-ray images, including the JSRT [125], OAI, NIH ChestXray14 [126], and NLST [109]. However, compared to MRI and CT, X-ray image registration investigations are few and far between [127, 128].

*3) MRI registration*

With a specific emphasis on brain MRIs, MRI [144, 148, 152] is the most widely used modality for developing image registration techniques due to the presence of large-scale public datasets (an example of cardiac and brain MRI registration is depicted in Figure 6). Thus, the most recently developed DLR techniques are verified against earlier state-of-the-art DLIR approaches, such as Conv2warp [115], VTN [57], and Voxelmorph [14, 103], in order to compare performance. The T1W and T2W modalities in most datasets on brain MRI are similarly used to build multi-modal medical image registration algorithms [26, 129]. Two publicly accessible datasets for cardiac motion estimation, Automatic Cardiac Diagnosis and Sunnybrook Cardiac Data Challenge, are available for cine MRI, which is the major modality utilized for cardiac image registration and cardiac motion estimation.

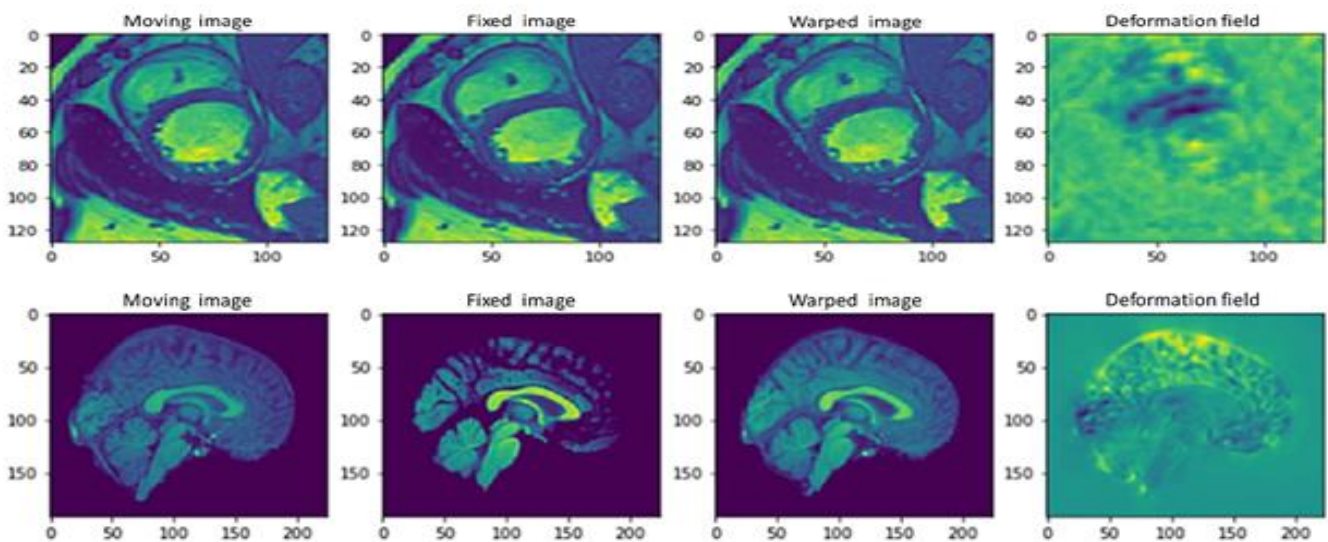

**Figure 6.** Voxelmorph-diff [104, 105] is an example of a cardiac and brain image registration. The first raw is cardiac MRI and second row is brain MRI registration [104, 105].

## B. Multi-modal registration

Medical image registration is a difficult challenge because of the lack of distinguishing characteristics, the non-rigid nature of the images, and the need for sub-pixel precision. Because of its capacity to compare pictures from diverse sources, DL has been frequently used for multi-modal registrations. However, unlike monomodal registration, there is a lack of open access datasets for multi-modal registration. When it comes to multi-modal registration, there are many researchers that gather and analyze their own datasets to create and test their algorithms. Several example multi-modal registration applications are discussed in this section: This includes 2D-3D registration, CT-MRI, CT-CBCCT, 2D-2W, and T1W-2W.

*1) 2D–3D registration*

Fixed and moving pictures in most applications for multi-modal registration explored so far have the same dimensions. Moreover, there are 3D picture volumes that can be used for 2D–2D image registration in public datasets. As a result, research has mostly concentrated on picture registration in 2D–2D and 3D–3D formats. Apart from these, there are several clinical uses for 2D–3D image registration, which is a significant component of current research towards multi-modal and DL-based image registration. Even more challenging is this endeavor since 2D images, such as those from x-rays, include tissue overlap and contrast difficulties inherent to 2D images. There have been a number of investigations into the 2D–3D image registration of X-ray to other 3D modality images, such as CT and MRI. There has also been considerable interest in slice-to-volume registration recently [130, 150]. DL-based 3D registration may be divided into 3 groups depending on the kind of DL technique as well as the training regime: deep iterative, supervised, and unsupervised. Fig. 7 depicts an overview of various strategies. Traditional iterative registration techniques are used in conjunction with deep iterative registration to include DL models. In this case, intensity-based similarity measurements are often substituted with deep similarity metrics. Deep similarity-based registration (DSBR) and RL-based registration are two subcategories of deep iterative registration. Different forms of reference data may be used to train DL models in supervised registration. These models may be more categorized as completely supervised, dual supervised or weakly supervised registration, depending on the kind of reference data employed. Likelihood GAN-based and metric-based registration are two subcategories of DL models that may be further categorized under unsupervised registration. In GAN-based registration, the discriminator and generator fight against each other in an adversarial way.

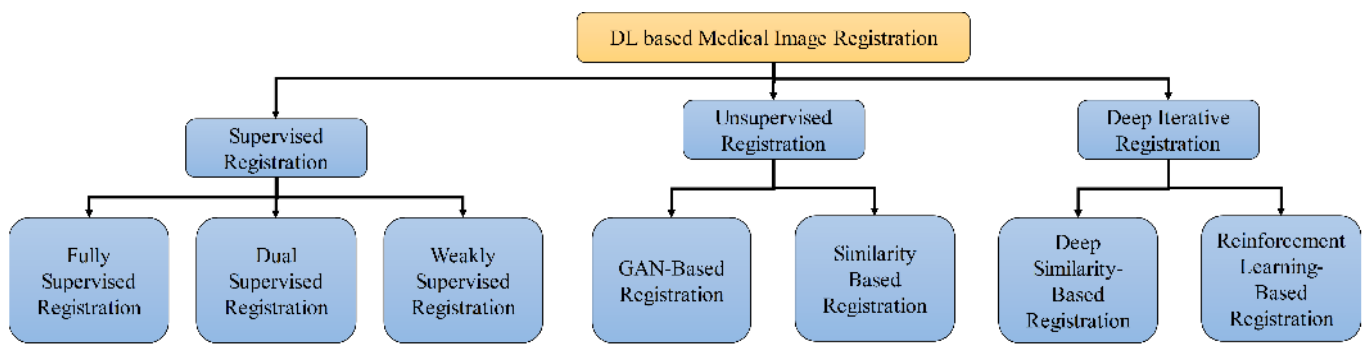

**Figure 7.** Various techniques of image registration

*2) CT-MRI registration*

Another popular multi-modal registration application is CT-MRI matching. 3 open access datasets for developing multi-modal registration techniques, including the three datasets discussed earlier, all comprise both CT and MRI scans of the same participants. The FFD structural representations on the RIRE dataset were learned using a PCANet developed by Zhu et al. [131]. According to Cao et al. [24], they developed an algorithm that could be used to match up MRI and CT images. Additionally, GAN-based networks have been used for pelvic imaging [30], while further research has presented methods for registering cardiac MRI and CT images [132].

*3) CT-CBCT registration*

Image registration in medical images between CT and CBCT images has received attention recently as well [54, 133]. The authors of [134] presented DCIGN to learn hierarchical features from the deformable registration of CT-CBCT on neck and head images. First, Yao [133] presented a CNN to predict a primary rough transformation, after that a typical intensity-based registration to improve the registration for image-guided radiotherapy (IGRT). This reduced the forecast time while maintaining high accuracy in the registrations.

*4) MRI-TRUS registration*

MRI and TRUS [151] images have also been compared in a number of studies. This endeavor requires the use of two publicly accessible datasets, BITE [135] and RESECT [136]. Several approaches have been created based on these datasets. Most of this research, on the other hand, is based on private information. To deal with inflexible registration of MRI-TRUS on prostate images, the authors of [31] developed a supervised network. Two networks, one for affine registration and one for deformable registration, were suggested by Hu and colleagues [17, 137]. To deal with this problem, Yan et al. [50] developed an adversarial image registration network (AIR-Net) based on GANs. An algorithm known as a neural network (NN) was used by Haskins and colleagues [66] to compare MRI to TRUS images.

*5) T1W-T2W registration*

With T1W to T2W registration, a mapping from T1-weighted to T2-weighted MRI images may be learned and applied in real time. Many freely accessible brain MRI datasets may be used for multi-modal image registration. On the basis of the IBIS 3D Autism Brain Imaging dataset, the authors of [138] developed a 3D Bayesian encoder to decoder network for multi-modal image registration. A GAN-based network, UMDIR, was suggested by Qin et al. [139] for this objective based on the BraTS 2017 dataset. The authors of [129] synthesized T1W. FLAIR, T2W, and T1 contrast-enhanced using a Cycle-GAN, providing a link across the various imaging techniques.

## VII. Challenges

The absence of training datasets with known transformations is one of the most significant difficulties for supervised DL-based techniques [141-147]. Various data augmentation techniques might be used to address this issue. Additional mistakes might be introduced by data augmentation techniques such as biases in arbitrary picture manipulation or changes between training and testing phases. There are many examples of organizations showing that the trained network can be used on datasets that are not the same as the training datasets. This prompted us to consider the possibility of using transfer learning to address the issue of insufficient training data. Transfer learning has not been used in the registration of medical images, which is surprising. In order to confine the expected transformation for unsupervised algorithms, multiple types of regularization terms were combined. However, determining the proportional value of each regularization term is a tough task. Researchers are still searching for the ideal selection of transformation regularization terms that may assist in building a deformation field for a certain registration job that is both physically feasible and physiologically realistic. In part, this is due to the absence of registration validation techniques. A lack of ground truth transformations between an image pair makes it almost impossible to compare registration approaches' results. As a result, registration validation techniques are just as crucial as the registration methods themselves. In 2019, there has been an upsurge in the number of articles that concentrate on registration validity. Further research into registration validation methods is required in order to accurately assess the efficacy of various registration techniques under various parametric settings.

## VIII. Future Scopes

Here, we present potential forthcoming research areas in DLIR to solve the issues raised up to now. An important first step is to understand what DLIR all is about. All registration approaches strive for accuracy, robustness, and speed. When trained to anticipate the spatial transformation matching the set of photos or pair, there was no substantial dissimilarity in registration time across techniques. Consequently, future DLIR techniques must put an emphasis on enhancing the networks' accuracy and generalization capabilities as well as on confirming that the predicted deformation fields are smooth and realistic.

1) Traditional approaches may be used with DL networks in a novel way. No matter how much faster and more accurate DLIR techniques have become, the advantages of traditional techniques (such as diffeomorphic characteristics and robust registration) cannot be ignored. DL networks and classical diffeomorphic transformations are being used to smooth out deformation fields.

2) As stated above, medical image registration is quite different from other medical image analysis jobs. DL networks may be made more application-specific by including additional picture registration prior to DL networks in future studies. To enhance the efficiency of registration, the morphology and topology of anatomical configurations, as well as the predicted kinds of deformation and spatial links between them, might all be included in DLIR networks. For example, additional labels might serve as the ground truth to assist the training process, even if ground-truth spatial transformations are occasionally accessible. Weakly-supervised medical image registration algorithms have been presented, and they typically outperform their unsupervised counterparts in terms of performance. Priors that are more informative when paired with synthetically changed training data, i.e., blackening pixels in the moving picture or producing adversarial instances [140], might improve the capacity of networks to generalize to unknown data while remaining resistant to varying image quality. For this reason, DL networks and spatial and temporal priors are potential study topics for the future.

Machine learning and computer vision are both advancing technologically, though, and new approaches are being proposed all the time. Neuroscience and machine learning may be brought together via the use of biologically realistic neuron models for computation in spiking neural networks (SNNs). In contrast, conventional neural networks are known to the machine learning community as SNNs. Instead of utilizing continuous values, SNNs use spikes, which are discrete events that occur at certain points in time. Differential equations reflect several biological processes, with the membrane potential being the most essential. As soon as a neuron reaches a certain potential, it spikes, resetting the neuron's potential to its original value. The Leaky Integrate-and-Fire (LIF) model is the most frequent. SNNs, on the other hand, are generally sparsely linked and use sparse network topologies to their advantage.

## IX. Conclusions

Recent efforts using DL for medical image registration have been reviewed in this article. It is necessary to develop the DL frameworks with care since each application has different problems. Multimodal image registrations, such as those involving TRUS and MRI, face similar challenges, such as the inability to use a robust similarity metric for multimodal applications, the lack of large datasets, the difficulty in obtaining ground truth registrations and segmentations, as well as quantifying the model's predilections. [36, 37]. Popular methods to these problems include patchwise frameworks, application-specific similarity metrics, registration frameworks and unsupervised techniques influenced by variational autoencoders. Interpolation and resampling, despite the intricacy of many of the approaches described in this paper, are typically not learnt by the neural network. As the area matures, we anticipate more academics to include these components into their DL-based solutions. Each strategy has its own merits and limitations, but the total number of researchers comparing the two is about equal. In both areas, we anticipate more studies and new approaches that combine the benefits of both tactics to emerge. We predict further studies in both categories.

## Acknowledgement

A part of this work was supported by the ICT innovation grant from the ICT Division of the Ministry of Posts, Telecommunications & Information Technology of the Government of the people's Republic of Bangladesh under Grant Code 1280101-120008431-3631108, GO-09, 2021-2022.

## References

[1] D. L. G. Hill, P. G. Batchelor, M. Holden, and D. J. Hawkes, "Medical image registration," Physics in medicine & biology, vol. 46, no. 3, p. R1, 2001.

[2] B. Zitova and J. Flusser, "Image registration methods: a survey," Image and vision computing, vol. 21, no. 11, pp. 977-1000, 2003.

[3] E. P. Ambinder, "A history of the shift toward full computerization of medicine," Journal of oncology practice, vol. 1, no. 2, p. 54, 2005.

[4] M. Z. Alom et al., "The history began from alexnet: A comprehensive survey on deep learning approaches," arXiv preprint arXiv:1803.01164, 2018.

[5] O. Ronneberger, P. Fischer, and T. Brox, "U-net: Convolutional networks for biomedical image segmentation," in International Conference on Medical image computing and computer-assisted intervention, 2015, pp. 234-241: Springer.

[6] K. He, X. Zhang, S. Ren, and J. Sun, "Deep residual learning for image recognition," in Proceedings of the IEEE conference on computer vision and pattern recognition, 2016, pp. 770-778.

[7] P. Podder, S. Bharati, and M. R. H. Mondal, "10 Automated gastric cancer detection and classification using machine learning," in Artificial Intelligence for Data-Driven Medical Diagnosis: De Gruyter, 2021, pp. 207-224.

[8] M. R. H. Mondal, S. Bharati, and P. Podder, "Diagnosis of COVID-19 Using Machine Learning and Deep Learning: A Review," Current Medical Imaging, 2021.

[9] S. Ren, K. He, R. Girshick, and J. Sun, "Faster r-cnn: Towards real-time object detection with region proposal networks," Advances in neural information processing systems, vol. 28, 2015.

[10] R. Liao et al., "An artificial agent for robust image registration," in Proceedings of the AAAI conference on artificial intelligence, 2017, vol. 31.

[11] K. Ma et al., "Multimodal image registration with deep context reinforcement learning," in International conference on medical image computing and computer-assisted intervention, 2017, pp. 240-248: Springer.

[12] S. Miao et al., "Dilated FCN for multi-agent 2D/3D medical image registration," in Proceedings of the AAAI Conference on Artificial Intelligence, 2018, vol. 32.

[13] B. D. De Vos, F. F. Berendsen, M. A. Viergever, H. Sokooti, M. Staring, and I. Išgum, "A deep learning framework for unsupervised affine and deformable image registration," Medical image analysis, vol. 52, pp. 128-143, 2019.

[14] H. Li and Y. Fan, "Non-rigid image registration using self-supervised fully convolutional networks without training data," in 2018 IEEE 15th International Symposium on Biomedical Imaging (ISBI 2018), 2018, pp. 1075-1078: IEEE.

[15] M. P. Heinrich et al., "MIND: Modality independent neighbourhood descriptor for multi-modal deformable registration," Medical image analysis, vol. 16, no. 7, pp. 1423-1435, 2012.

[16] J. Fan, X. Cao, Z. Xue, P.-T. Yap, and D. Shen, "Adversarial similarity network for evaluating image alignment in deep learning based registration," in International Conference on Medical Image Computing and Computer-Assisted Intervention, 2018, pp. 739-746: Springer.

[17] Y. Hu et al., "Weakly-supervised convolutional neural networks for multimodal image registration," Medical image analysis, vol. 49, pp. 1-13, 2018.

[18] F. P. M. Oliveira and J. M. R. S. Tavares, "Medical image registration: a review," Computer methods in biomechanics and biomedical engineering, vol. 17, no. 2, pp. 73-93, 2014.

[19] A. Paszke et al., "Automatic differentiation in pytorch," 2017.

[20] Y. Jia et al., "Caffe: Convolutional architecture for fast feature embedding," in Proceedings of the 22nd ACM international conference on Multimedia, 2014, pp. 675-678.

[21] N. Ketkar, "Introduction to keras," in Deep learning with Python: Springer, 2017, pp. 97-111.

[22] T. Chen et al., "Mxnet: A flexible and efficient machine learning library for heterogeneous distributed systems," arXiv preprint arXiv:1512.01274, 2015.

[23] M. Abadi et al., "{TensorFlow}: A System for {Large-Scale} Machine Learning," in 12th USENIX symposium on operating systems design and implementation (OSDI 16), 2016, pp. 265-283.

[24] X. Cao, J. Yang, L. Wang, Z. Xue, Q. Wang, and D. Shen, "Deep learning based inter-modality image registration supervised by intra-modality similarity," in International workshop on machine learning in medical imaging, 2018, pp. 55-63: Springer.

[25] X. Teng, Y. Chen, Y. Zhang, and L. Ren, "Respiratory deformation registration in 4D-CT/cone beam CT using deep learning," Quantitative Imaging in Medicine and Surgery, vol. 11, no. 2, p. 737, 2021.

[26] X. Yang, R. Kwitt, M. Styner, and M. Niethammer, "Quicksilver: Fast predictive image registration–a deep learning approach," NeuroImage, vol. 158, pp. 378-396, 2017.

[27] M.-M. Rohé, M. Datar, T. Heimann, M. Sermesant, and X. Pennec, "SVF-Net: learning deformable image registration using shape matching," in International conference on medical image computing and computer-assisted intervention, 2017, pp. 266-274: Springer.

[28] J. Wang and M. Zhang, "Deep Learning for Regularization Prediction in Diffeomorphic Image Registration," arXiv preprint arXiv:2011.14229, 2020.

[29] H. Sokooti et al., "3D convolutional neural networks image registration based on efficient supervised learning from artificial deformations," arXiv preprint arXiv:1908.10235, 2019.

[30] H. Sokooti, B. d. Vos, F. Berendsen, B. P. F. Lelieveldt, I. Išgum, and M. Staring, "Nonrigid image registration using multi-scale 3D convolutional neural networks," in International conference on medical image computing and computer-assisted intervention, 2017, pp. 232-239: Springer.

[31] H. Guo, M. Kruger, S. Xu, B. J. Wood, and P. Yan, "Deep adaptive registration of multi-modal prostate images," Computerized Medical Imaging and Graphics, vol. 84, p. 101769, 2020.

[32] Y. Fu et al., "Synthetic CT-aided MRI-CT image registration for head and neck radiotherapy," in Medical Imaging 2020: Biomedical Applications in Molecular, Structural, and Functional Imaging, 2020, vol. 11317, p. 1131728: International Society for Optics and Photonics.

[33] J. Fan, X. Cao, P.-T. Yap, and D. Shen, "BIRNet: Brain image registration using dual-supervised fully convolutional networks," Medical image analysis, vol. 54, pp. 193-206, 2019.

[34] S. Ahmad, J. Fan, P. Dong, X. Cao, P.-T. Yap, and D. Shen, "Deep learning deformation initialization for rapid groupwise registration of inhomogeneous image populations," Frontiers in Neuroinformatics, vol. 13, p. 34, 2019.

[35] I. Y. Ha, L. Hansen, M. Wilms, and M. P. Heinrich, "Geometric deep learning and heatmap prediction for large deformation registration of abdominal and thoracic CT," in International Conference on Medical Imaging with Deep Learning--Extended Abstract Track, 2019.

[36] T. Estienne et al., "Deep learning based registration using spatial gradients and noisy segmentation labels," in International Conference on Medical Image Computing and Computer-Assisted Intervention, 2020, pp. 87-93: Springer.

[37] A. Hering, B. v. Ginneken, and S. Heldmann, "mlvirnet: Multilevel variational image registration network," in International Conference on Medical Image Computing and Computer-Assisted Intervention, 2019, pp. 257-265: Springer.

[38] S. Hu, L. Zhang, G. Li, M. Liu, D. Fu, and W. Zhang, "Infant brain deformable registration using global and local label-driven deep regression learning," in International Workshop on Machine Learning in Medical Imaging, 2019, pp. 106-114: Springer.

[39] B. Li et al., "A hybrid deep learning framework for integrated segmentation and registration: evaluation on longitudinal white matter tract changes," in International Conference on Medical Image Computing and Computer-Assisted Intervention, 2019, pp. 645-653: Springer.

[40] Z. Xu and M. Niethammer, "DeepAtlas: Joint semi-supervised learning of image registration and segmentation," in International Conference on Medical Image Computing and Computer-Assisted Intervention, 2019, pp. 420-429: Springer.

[41] G. Balakrishnan, A. Zhao, M. R. Sabuncu, J. Guttag, and A. V. Dalca, "VoxelMorph: a learning framework for deformable medical image registration," IEEE transactions on medical imaging, vol. 38, no. 8, pp. 1788-1800, 2019.

[42] Z. Zhu et al., "Joint affine and deformable three‐dimensional networks for brain MRI registration," Medical Physics, vol. 48, no. 3, pp. 1182-1196, 2021.

[43] S. Bharati, P. Podder, and M. R. H. Mondal, "Artificial Neural Network Based Breast Cancer Screening: A Comprehensive Review," International Journal of Computer Information Systems and Industrial Management Applications, vol. 12, pp. 125-137, 2020.

[44] S. Bharati, M. A. Rahman, and P. Podder, "Breast cancer prediction applying different classification algorithm with comparative analysis using WEKA," in 2018 4th International Conference on Electrical Engineering and Information & Communication Technology (iCEEiCT), Dhaka, Bangladesh, 2018, pp. 581-584: IEEE.

[45] S. Bharati, T. Z. Khan, P. Podder, and N. Q. Hung, "A Comparative Analysis of Image Denoising Problem: Noise Models, Denoising Filters and Applications," Cognitive Internet of Medical Things for Smart Healthcare, 2020.

[46] P. Podder, A. Khamparia, M. R. H. Mondal, M. A. Rahman, and S. Bharati, "Forecasting the Spread of COVID-19 and ICU Requirements," International Journal of Online and Biomedical Engineering (iJOE), no. 5, 2021.

[47] S. Bharati, P. Podder, and M. R. H. Mondal, "Hybrid deep learning for detecting lung diseases from X-ray images," Informatics in Medicine Unlocked, p. 100391, 2020.

[48] T. Estienne et al., "U-ReSNet: Ultimate coupling of registration and segmentation with deep nets," in International conference on medical image computing and computer-assisted intervention, 2019, pp. 310-319: Springer.

[49] B. Kim, J. Kim, J.-G. Lee, D. H. Kim, S. H. Park, and J. C. Ye, "Unsupervised deformable image registration using cycle-consistent cnn," in International Conference on Medical Image Computing and Computer-Assisted Intervention, 2019, pp. 166-174: Springer.

[50] P. Yan, S. Xu, A. R. Rastinehad, and B. J. Wood, "Adversarial image registration with application for MR and TRUS image fusion," in International Workshop on Machine Learning in Medical Imaging, 2018, pp. 197-204: Springer.

[51] Y. Fu et al., "Biomechanically constrained non-rigid MR-TRUS prostate registration using deep learning based 3D point cloud matching," Medical image analysis, vol. 67, p. 101845, 2021.

[52] Y. Fu et al., "LungRegNet: An unsupervised deformable image registration method for 4D‐CT lung," Medical physics, vol. 47, no. 4, pp. 1763-1774, 2020.

[53] D. Kuang, "Cycle-Consistent training for reducing negative Jacobian determinant in deep registration networks," in International Workshop on Simulation and Synthesis in Medical Imaging, 2019, pp. 120-129: Springer.

[54] Z. Jiang, F.-F. Yin, Y. Ge, and L. Ren, "A multi-scale framework with unsupervised joint training of convolutional neural networks for pulmonary deformable image registration," Physics in Medicine & Biology, vol. 65, no. 1, p. 015011, 2020.

[55] Z. Shen, X. Han, Z. Xu, and M. Niethammer, "Networks for joint affine and non-parametric image registration," in Proceedings of the IEEE/CVF Conference on Computer Vision and Pattern Recognition, 2019, pp. 4224-4233.

[56] H. Yu et al., "Learning 3D non-rigid deformation based on an unsupervised deep learning for PET/CT image registration," in Medical Imaging 2019: Biomedical Applications in Molecular, Structural, and Functional Imaging, 2019, vol. 10953, pp. 439-444: SPIE.

[57] S. Zhao, T. Lau, J. Luo, I. Eric, C. Chang, and Y. Xu, "Unsupervised 3D end-to-end medical image registration with volume tweening network," IEEE journal of biomedical and health informatics, vol. 24, no. 5, pp. 1394-1404, 2019.

[58] Y. Guo, X. Wu, Z. Wang, X. Pei, and X. G. Xu, "End‐to‐end unsupervised cycle‐consistent fully convolutional network for 3D pelvic CT‐MR deformable registration," Journal of Applied Clinical Medical Physics, vol. 21, no. 9, pp. 193-200, 2020.

[59] Y. Lei et al., "4D-CT deformable image registration using multiscale unsupervised deep learning," Physics in Medicine & Biology, vol. 65, no. 8, p. 085003, 2020.

[60] W. Shao et al., "ProsRegNet: A deep learning framework for registration of MRI and histopathology images of the prostate," Medical image analysis, vol. 68, p. 101919, 2021.

[61] L. Duan et al., "Unsupervised learning for deformable registration of thoracic CT and cone‐beam CT based on multiscale features matching with spatially adaptive weighting," Medical Physics, vol. 47, no. 11, pp. 5632-5647, 2020.

[62] M. S. Elmahdy et al., "Robust contour propagation using deep learning and image registration for online adaptive proton therapy of prostate cancer," Medical physics, vol. 46, no. 8, pp. 3329-3343, 2019.

[63] J. Fan, X. Cao, Q. Wang, P.-T. Yap, and D. Shen, "Adversarial learning for mono-or multi-modal registration," Medical image analysis, vol. 58, p. 101545, 2019.

[64] Y. Huang, S. Ahmad, J. Fan, D. Shen, and P.-T. Yap, "Difficulty-aware hierarchical convolutional neural networks for deformable registration of brain mr images," Medical image analysis, vol. 67, p. 101817, 2021.

[65] Y. Fu, Y. Lei, T. Wang, W. J. Curran, T. Liu, and X. Yang, "Deep learning in medical image registration: a review," Physics in Medicine & Biology, vol. 65, no. 20, p. 20TR01, 2020.

[66] G. Haskins, U. Kruger, and P. Yan, "Deep learning in medical image registration: a survey," Machine Vision and Applications, vol. 31, no. 1, pp. 1-18, 2020.

[67] F. Maes, A. Collignon, D. Vandermeulen, G. Marchal, and P. Suetens, "Multimodality image registration by maximization of mutual information," IEEE transactions on Medical Imaging, vol. 16, no. 2, pp. 187-198, 1997.

[68] P. Viola and W. M. Wells Iii, "Alignment by maximization of mutual information," International journal of computer vision, vol. 24, no. 2, pp. 137-154, 1997.

[69] M. Simonovsky, B. Gutiérrez-Becker, D. Mateus, N. Navab, and N. Komodakis, "A deep metric for multimodal registration," in International conference on medical image computing and computer-assisted intervention, 2016, pp. 10-18: Springer.

[70] G. Wu, M. Kim, Q. Wang, Y. Gao, S. Liao, and D. Shen, "Unsupervised deep feature learning for deformable registration of MR brain images," in International Conference on Medical Image Computing and Computer-Assisted Intervention, 2013, pp. 649-656: Springer.

[71] G. Wu, M. Kim, Q. Wang, B. C. Munsell, and D. Shen, "Scalable high-performance image registration framework by unsupervised deep feature representations learning," IEEE Transactions on Biomedical Engineering, vol. 63, no. 7, pp. 1505-1516, 2015.

[72] J. Krebs et al., "Robust non-rigid registration through agent-based action learning," in International Conference on Medical Image Computing and Computer-Assisted Intervention, 2017, pp. 344-352: Springer.

[73] M. Blendowski and M. P. Heinrich, "Combining MRF-based deformable registration and deep binary 3D-CNN descriptors for large lung motion estimation in COPD patients," International journal of computer assisted radiology and surgery, vol. 14, no. 1, pp. 43-52, 2019.

[74] X. Cheng, L. Zhang, and Y. Zheng, "Deep similarity learning for multimodal medical images," Computer Methods in Biomechanics and Biomedical Engineering: Imaging & Visualization, vol. 6, no. 3, pp. 248-252, 2018.

[75] K. A. J. Eppenhof and J. P. W. Pluim, "Error estimation of deformable image registration of pulmonary CT scans using convolutional neural networks," Journal of Medical Imaging, vol. 5, no. 2, p. 024003, 2018.

[76] R. Wright et al., "LSTM spatial co-transformer networks for registration of 3D fetal US and MR brain images," in Data Driven Treatment Response Assessment and Preterm, Perinatal, and Paediatric Image Analysis: Springer, 2018, pp. 149-159.

[77] A. Sedghi et al., "Semi-supervised deep metrics for image registration," arXiv preprint arXiv:1804.01565, 2018.

[78] D. Shen, "Image registration by local histogram matching," Pattern Recognition, vol. 40, no. 4, pp. 1161-1172, 2007.

[79] G. Haskins et al., "Learning deep similarity metric for 3D MR–TRUS image registration," International journal of computer assisted radiology and surgery, vol. 14, no. 3, pp. 417-425, 2019.

[80] A. Sedghi, L. J. O'Donnell, T. Kapur, E. Learned-Miller, P. Mousavi, and W. M. Wells Iii, "Image registration: Maximum likelihood, minimum entropy and deep learning," Medical Image Analysis, vol. 69, p. 101939, 2021.

[81] S. Czolbe, O. Krause, and A. Feragen, "DeepSim: Semantic similarity metrics for learned image registration," arXiv preprint arXiv:2011.05735, 2020.

[82] R. W. K. So and A. C. S. Chung, "A novel learning-based dissimilarity metric for rigid and non-rigid medical image registration by using Bhattacharyya Distances," Pattern Recognition, vol. 62, pp. 161-174, 2017.

[83] M. Niethammer, R. Kwitt, and F.-X. Vialard, "Metric learning for image registration," in Proceedings of the IEEE/CVF Conference on Computer Vision and Pattern Recognition, 2019, pp. 8463-8472.

[84] Z. Wang, T. Schaul, M. Hessel, H. Hasselt, M. Lanctot, and N. Freitas, "Dueling network architectures for deep reinforcement learning," in International conference on machine learning, 2016, pp. 1995-2003: PMLR.

[85] T. De Silva et al., "3D–2D image registration for target localization in spine surgery: investigation of similarity metrics providing robustness to content mismatch," Physics in Medicine & Biology, vol. 61, no. 8, p. 3009, 2016.

[86] M. Lorenzi, N. Ayache, G. B. Frisoni, X. Pennec, and I. Alzheimer's Disease Neuroimaging, "LCC-Demons: a robust and accurate symmetric diffeomorphic registration algorithm," NeuroImage, vol. 81, pp. 470-483, 2013.

[87] S. Klein, M. Staring, K. Murphy, M. A. Viergever, and J. P. W. Pluim, "Elastix: a toolbox for intensity-based medical image registration," IEEE transactions on medical imaging, vol. 29, no. 1, pp. 196-205, 2009.

[88] J. Hu et al., "End-to-end multimodal image registration via reinforcement learning," Medical Image Analysis, vol. 68, p. 101878, 2021.

[89] A. Dalca, M. Rakic, J. Guttag, and M. Sabuncu, "Learning conditional deformable templates with convolutional networks," Advances in neural information processing systems, vol. 32, 2019.

[90] A. V. Dalca, E. Yu, P. Golland, B. Fischl, M. R. Sabuncu, and J. Eugenio Iglesias, "Unsupervised deep learning for Bayesian brain MRI segmentation," in International Conference on Medical Image Computing and Computer-Assisted Intervention, 2019, pp. 356-365: Springer.

[91] P. Podder, S. Bharati, M. A. Rahman, and U. Kose, "Transfer Learning for Classification of Brain Tumor," in Deep Learning for Biomedical Applications: CRC Press, 2021, pp. 315-328.

[92] P. Radau, Y. Lu, K. Connelly, G. Paul, A. Dick, and G. Wright, "Evaluation framework for algorithms segmenting short axis cardiac MRI," The MIDAS Journal-Cardiac MR Left Ventricle Segmentation Challenge, vol. 49, 2009.

[93] B. D. d. Vos, F. F. Berendsen, M. A. Viergever, M. Staring, and I. Išgum, "End-to-end unsupervised deformable image registration with a convolutional neural network," in Deep learning in medical image analysis and multimodal learning for clinical decision support: Springer, 2017, pp. 204-212.

[94] Y. Sang and D. Ruan, "Enhanced image registration with a network paradigm and incorporation of a deformation representation model," in 2020 IEEE 17th International Symposium on Biomedical Imaging (ISBI), 2020, pp. 91-94: IEEE.

[95] T. Fechter and D. Baltas, "One-shot learning for deformable medical image registration and periodic motion tracking," IEEE transactions on medical imaging, vol. 39, no. 7, pp. 2506-2517, 2020.

[96] A. Agustí et al., "Persistent systemic inflammation is associated with poor clinical outcomes in COPD: a novel phenotype," PloS one, vol. 7, no. 5, p. e37483, 2012.

[97] S. Zhao, Y. Dong, E. I. Chang, and Y. Xu, "Recursive cascaded networks for unsupervised medical image registration," in Proceedings of the IEEE/CVF International Conference on Computer Vision, 2019, pp. 10600-10610.

[98] A. Dagley et al., "Harvard aging brain study: dataset and accessibility," Neuroimage, vol. 144, pp. 255-258, 2017.

[99] R. L. Gollub et al., "The MCIC collection: a shared repository of multi-modal, multi-site brain image data from a clinical investigation of schizophrenia," Neuroinformatics, vol. 11, no. 3, pp. 367-388, 2013.

[100] S. G. Mueller et al., "Ways toward an early diagnosis in Alzheimer's disease: the Alzheimer's Disease Neuroimaging Initiative (ADNI)," Alzheimer's & Dementia, vol. 1, no. 1, pp. 55-66, 2005.

[101] K. Marek et al., "The Parkinson progression marker initiative (PPMI)," Progress in neurobiology, vol. 95, no. 4, pp. 629-635, 2011.

[102] [102] A. Di Martino et al., "The autism brain imaging data exchange: towards a large-scale evaluation of the intrinsic brain architecture in autism," Molecular psychiatry, vol. 19, no. 6, pp. 659-667, 2014.

[103] G. Balakrishnan, A. Zhao, M. R. Sabuncu, J. Guttag, and A. V. Dalca, "An unsupervised learning model for deformable medical image registration," in Proceedings of the IEEE conference on computer vision and pattern recognition, 2018, pp. 9252-9260.

[104] A. V. Dalca, G. Balakrishnan, J. Guttag, and M. R. Sabuncu, "Unsupervised learning for fast probabilistic diffeomorphic registration," in International Conference on Medical Image Computing and Computer-Assisted Intervention, 2018, pp. 729-738: Springer.

[105] A. V. Dalca, G. Balakrishnan, J. Guttag, and M. R. Sabuncu, "Unsupervised learning of probabilistic diffeomorphic registration for images and surfaces," Medical image analysis, vol. 57, pp. 226-236, 2019.

[106] Z. Shen, F.-X. Vialard, and M. Niethammer, "Region-specific diffeomorphic metric mapping," Advances in Neural Information Processing Systems, vol. 32, 2019.

[107] R. Castillo et al., "A framework for evaluation of deformable image registration spatial accuracy using large landmark point sets," Physics in Medicine & Biology, vol. 54, no. 7, p. 1849, 2009.

[108] R. Castillo et al., "A reference dataset for deformable image registration spatial accuracy evaluation using the COPDgene study archive," Physics in Medicine & Biology, vol. 58, no. 9, p. 2861, 2013.

[109] T. National Lung Screening Trial Research, "Reduced lung-cancer mortality with low-dose computed tomographic screening," New England Journal of Medicine, vol. 365, no. 5, pp. 395-409, 2011.

[110] E. A. Regan et al., "Genetic epidemiology of COPD (COPDGene) study design," COPD: Journal of Chronic Obstructive Pulmonary Disease, vol. 7, no. 1, pp. 32-43, 2011.

[111] J. Vandemeulebroucke, S. Rit, J. Kybic, P. Clarysse, and D. Sarrut, "Spatiotemporal motion estimation for respiratory‐correlated imaging of the lungs," Medical physics, vol. 38, no. 1, pp. 166-178, 2011.

[112] S. G. Armato Iii et al., "The lung image database consortium (LIDC) and image database resource initiative (IDRI): a completed reference database of lung nodules on CT scans," Medical physics, vol. 38, no. 2, pp. 915-931, 2011.

[113] K. A. J. Eppenhof and J. P. W. Pluim, "Pulmonary CT registration through supervised learning with convolutional neural networks," IEEE transactions on medical imaging, vol. 38, no. 5, pp. 1097-1105, 2018.

[114] K. A. J. Eppenhof, M. W. Lafarge, M. Veta, and J. P. W. Pluim, "Progressively trained convolutional neural networks for deformable image registration," IEEE transactions on medical imaging, vol. 39, no. 5, pp. 1594-1604, 2019.

[115] S. Ali and J. Rittscher, "Conv2warp: An unsupervised deformable image registration with continuous convolution and warping," pp. 489-497: Springer.

[116] S. Bharati and P. Podder, "1 Performance of CNN for predicting cancerous lung nodules using LightGBM," in

Artificial Intelligence for Data-Driven Medical Diagnosis: De Gruyter, 2021, pp. 1-18.
[117]   M. R. H. Mondal, S. Bharati, and P. Podder, "CO-IRv2: Optimized InceptionResNetV2 for COVID-19 detection from chest CT images," PloS one, vol. 16, no. 10, p. e0259179, 2021.
[118]   S. Bharati, P. Podder, R. Mondal, A. Mahmood, and M. Raihan-Al-Masud, "Comparative performance analysis of different classification algorithm for the purpose of prediction of lung cancer," in International Conference on Intelligent Systems Design and Applications, 2018, vol. 941, pp. 447-457: Springer.
[119]   S. Bharati, P. Podder, M. Mondal, and V. B. Prasath, "CO-ResNet: Optimized ResNet model for COVID-19 diagnosis from X-ray images," International Journal of Hybrid Intelligent Systems, vol. 17, pp. 71-85, 2021.
[120]   A. Khamparia et al., "Diagnosis of Breast Cancer Based on Modern Mammography using Hybrid Transfer Learning," Multidimensional Systems and Signal Processing, 2020.
[121]   S. Bharati, P. Podder, and P. K. Paul, "Lung cancer recognition and prediction according to random forest ensemble and RUSBoost algorithm using LIDC data," International Journal of Hybrid Intelligent Systems, vol. 15, no. 2, pp. 91-100, 2019.
[122]   S. Bharati, P. Podder, M. Mondal, and V. B. Prasath, "Medical Imaging with Deep Learning for COVID-19 Diagnosis: A Comprehensive Review," International Journal of Computer Information Systems and Industrial Management Applications, vol. 13, pp. 91 - 112, 2021.
[123]   S. Bharati, P. Podder, M. R. H. Mondal, and N. Gandhi, "Optimized NASNet for Diagnosis of COVID-19 from Lung CT Images," presented at the 20th International Conference on Intelligent Systems Design and Applications (ISDA 2020), 2020.
[124]   L. Canalini, J. Klein, D. Miller, and R. Kikinis, "Segmentation-based registration of ultrasound volumes for glioma resection in image-guided neurosurgery," International journal of computer assisted radiology and surgery, vol. 14, no. 10, pp. 1697-1713, 2019.
[125]   S. Junji et al., "Development of a digital image database for chest radiographs with and without a lung nodule: receiver operating characteristic analysis of radiologists' detection of pulmonary nodules," American Journal of Roentgenology, vol. 174, no. 1, pp. 71-74, 2000.
[126]   X. Wang, Y. Peng, L. Lu, Z. Lu, M. Bagheri, and R. M. Summers, "Chestx-ray8: Hospital-scale chest x-ray database and benchmarks on weakly-supervised classification and localization of common thorax diseases," in Proceedings of the IEEE conference on computer vision and pattern recognition, 2017, pp. 2097-2106.
[127]   D. Mahapatra and Z. Ge, "Training data independent image registration with gans using transfer learning and segmentation information," in 2019 IEEE 16th International Symposium on Biomedical Imaging (ISBI 2019), 2019, pp. 709-713: IEEE.
[128]   L. Mansilla, D. H. Milone, and E. Ferrante, "Learning deformable registration of medical images with anatomical constraints," Neural Networks, vol. 124, pp. 269-279, 2020.
[129]   Z. Tang, P.-T. Yap, and D. Shen, "A new multi-atlas registration framework for multimodal pathological images using conventional monomodal normal atlases," IEEE Transactions on Image Processing, vol. 28, no. 5, pp. 2293-2304, 2018.
[130]   S. S. M. Salehi, S. Khan, D. Erdogmus, and A. Gholipour, "Real-time deep pose estimation with geodesic loss for image-to-template rigid registration," IEEE transactions on medical imaging, vol. 38, no. 2, pp. 470-481, 2018.
[131]   X. Zhu, M. Ding, T. Huang, X. Jin, and X. Zhang, "PCANet-based structural representation for nonrigid multimodal medical image registration," Sensors, vol. 18, no. 5, p. 1477, 2018.
[132]   A. Hering, S. Kuckertz, S. Heldmann, and M. P. Heinrich, "Memory-efficient 2.5 D convolutional transformer networks for multi-modal deformable registration with weak label supervision applied to whole-heart CT and MRI scans," International journal of computer assisted radiology and surgery, vol. 14, no. 11, pp. 1901-1912, 2019.
[133]   Z. Yao et al., "A supervised network for fast image-guided radiotherapy (IGRT) registration," Journal of medical systems, vol. 43, no. 7, pp. 1-8, 2019.
[134]   V. Kearney, S. Haaf, A. Sudhyadhom, G. Valdes, and T. D. Solberg, "An unsupervised convolutional neural network-based algorithm for deformable image registration," Physics in Medicine & Biology, vol. 63, no. 18, p. 185017, 2018.
[135]   L. Mercier, R. F. Del Maestro, K. Petrecca, D. Araujo, C. Haegelen, and D. L. Collins, "Online database of clinical MR and ultrasound images of brain tumors," Medical physics, vol. 39, no. 6Part1, pp. 3253-3261, 2012.
[136]   Y. Xiao, M. Fortin, G. Unsgård, H. Rivaz, and I. Reinertsen, "RE troSpective Evaluation of Cerebral Tumors (RESECT): A clinical database of pre‐operative MRI and intra‐operative ultrasound in low‐grade glioma surgeries," Medical physics, vol. 44, no. 7, pp. 3875-3882, 2017.
[137]   Y. Hu et al., "Label-driven weakly-supervised learning for multimodal deformable image registration," in 2018 IEEE 15th International Symposium on Biomedical Imaging (ISBI 2018), 2018, pp. 1070-1074: IEEE.
[138]   X. Yang, R. Kwitt, M. Styner, and M. Niethammer, "Fast predictive multimodal image registration," in 2017 IEEE 14th International Symposium on Biomedical Imaging (ISBI 2017), 2017, pp. 858-862: IEEE.
[139]   C. Qin, B. Shi, R. Liao, T. Mansi, D. Rueckert, and A. Kamen, "Unsupervised deformable registration for multi-modal images via disentangled representations," in International Conference on Information Processing in Medical Imaging, 2019, pp. 249-261: Springer.
[140]   X. Ma et al., "Understanding adversarial attacks on deep learning based medical image analysis systems," Pattern Recognition, vol. 110, p. 107332, 2021.
[141]   S. Bharati, P. Podder, M. R. H. Mondal, V. B. S. Prasath, and N. Gandhi, "Ensemble Learning for Data-Driven Diagnosis of Polycystic Ovary Syndrome," presented at the 21st International Conference on Intelligent Systems Design and Applications (ISDA

2021), vol 418. Springer, Cham, 2021. https://doi.org/10.1007/978-3-030-96308-8_116

[142] A. M. Begum, M. R. H. Mondal, P. Podder, and S. Bharati, "Detecting Spinal Abnormalities using Multilayer Perceptron Algorithm," presented at the 11th World Congress on Information and Communication Technologies, pp. 654-664. Springer, Cham, 2021. https://doi.org/10.1007/978-3-030-96299-9_62

[143] S. Bharati, P. Podder, M. R. H. Mondal, P. Podder, and U. Kose, "A review on epidemiology, genomic characteristics, spread, and treatments of COVID-19," in Data Science for COVID-19: Elsevier, 2022, pp. 487-505.

[144] S. Bharati, P. Podder, D. N. H. Thanh, V. B. S. Prasath. "Dementia classification using MR imaging and voting based machine learning models", Multimedia Tools and Applications, 2022. doi:10.1007/s11042-022-12754-x

[145] P. Podder, S. Bharati, M. R. H. Mondal, and U. Kose, "Application of Machine Learning for the Diagnosis of COVID-19," in Data Science for COVID-19: Elsevier, 2021, pp. 175-194.

[146] S. Bharati, P. Podder, and M. R. H. Mondal. "Diagnosis of polycystic ovary syndrome using machine learning algorithms." In 2020 IEEE Region 10 Symposium (TENSYMP), pp. 1486-1489. IEEE, 2020.

[147] S. Bharati, P. Podder, M.R.H. Mondal, and P.K. Paul. "Applications and challenges of cloud integrated IoMT." In Cognitive Internet of Medical Things for Smart Healthcare, pp. 67-85. Springer, Cham, 2021.

[148] S. Bharati, P. Podder, and Md Raihan Al-Masud. "Brain Magnetic Resonance Imaging Compression Using Daubechies & Biorthogonal Wavelet with the Fusion of STW and SPIHT." In 2018 International Conference on Advancement in Electrical and Electronic Engineering (ICAEEE), pp. 1-4. IEEE, 2018.

[149] R. Ahmed, K. E. K. Emon, and M. F. Hossain. "Robust driver fatigue recognition using image processing." In 2014 International Conference on Informatics, Electronics & Vision (ICIEV), pp. 1-6. IEEE, 2014.

[150] N. Masoumi, C. J. Belasso, M. O. Ahmad, H. Benali, Y. Xiao, and H. Rivaz. "Multimodal 3D ultrasound and CT in image-guided spinal surgery: public database and new registration algorithms." International Journal of Computer Assisted Radiology and Surgery 16, no. 4 (2021): 555-565.

[151] A. Pirhadi, H. Rivaz, M. O. Ahmad, and Y. Xiao. "Robust Ultrasound-to-Ultrasound Registration for Intra-operative Brain Shift Correction with a Siamese Neural Network." In International Workshop on Advances in Simplifying Medical Ultrasound, pp. 85-95. Springer, Cham, 2021.

[152] R. Basnet, M. O. Ahmad, and M. N. S. Swamy. "A deep dense residual network with reduced parameters for volumetric brain tissue segmentation from MR images." Biomedical Signal Processing and Control 70 (2021): 103063.

[153] A. Esmaeilzehi, M. O. Ahmad, and M. N. S. Swamy. "SRNMSM: A Deep Light-Weight Image Super Resolution Network Using Multi-Scale Spatial and Morphological Feature Generating Residual Blocks." IEEE Transactions on Broadcasting (2021).

[154] S. Bharati, M. A. Rahman, S. Mandal, and P. Podder. "Analysis of DWT, DCT, BFO & PBFO algorithm for the purpose of medical image watermarking." In 2018 International Conference on Innovation in Engineering and Technology (ICIET), pp. 1-6. IEEE, 2018.

## Author Biographies

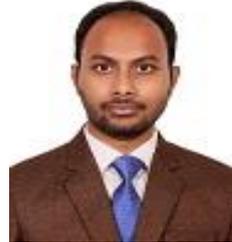

**Subrato Bharati** received a BS degree in Electrical and Electronic Engineering from Ranada Prasad Shaha University, Bangladesh. He is currently working as a researcher in the Institute of Information and Communication Technology, Bangladesh University of Engineering and Technology, Dhaka, Bangladesh. He is a regular reviewer of a number of international journals, including from IEEE, Elsevier, Springer, Wiley, and other reputed publishers. He is an associate editor of the Journal of the International Academy for Case Studies and guest editor of a special issue in the Journal of Internet Technology (SCI Index Journal). He has been a member of scientific and technical program committees at conferences such as CECNet 2021, ICONCS, ICCRDA 2020, ICICCR 2021, CECIT 2021, and others. His research interest includes bioinformatics, medical image processing, pattern recognition, deep learning, wireless communications, data analytics, machine learning, neural networks, and feature selection. He has published a number of papers in the journals of Elsevier, Springer, PLOS, IOS Press, and others and has published several IEEE and Springer reputed conference papers. He edited 3 books. He has also published Springer, Elsevier, De Gruyter, CRC Press and Wiley Book chapters.

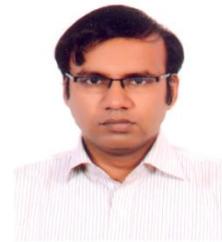

**M. Rubaiyat Hossain Mondal** received BSc and MSc degrees in electrical and electronic engineering from Bangladesh University of Engineering and Technology (BUET), Dhaka, Bangladesh. He obtained a PhD degree in 2014 from the Department of Electrical and Computer Systems Engineering, Monash University, Melbourne, Australia. From 2005 to 2010 and from 2014 to date, he has been working as a faculty member at the Institute of Information and Communication Technology (IICT) in BUET, Bangladesh. He has published a number of papers in journals of IEEE, IET, Elsevier, Springer, Wiley, De Gruyter, PLOS, MDPI, and more. He has published several conference papers and book chapters and edited a book published by De Gruyter in 2021. He has so far successfully supervised 10 students in completing their masters' theses in the field of Information and Communication Technology at BUET, Bangladesh. His research interests include artificial intelligence, image processing, bioinformatics, wireless communications and cryptography. For more information, https://rubaiyat97.buet.ac.bd/

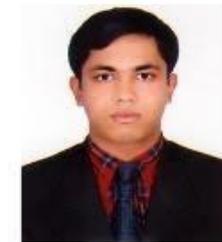

**Prajoy Podder** is currently a researcher at the Institute of Information and Communication Technology, Bangladesh University of Engineering and Technology, Dhaka, Bangladesh. He has also worked as a lecturer in the Department of Electrical and Electronic Engineering, Ranada Prasad Shaha University, Narayanganj, Bangladesh. He received a BSc (Engg) degree in Electronics and Communication Engineering from the Khulna University of Engineering and Technology in Khulna, Bangladesh, in 2014. He recently completed an MSc in Information and Communication Technology from the Bangladesh University of Engineering and Technology. He has authored or co-authored over 45 journal articles, conference proceedings and book chapters published by IEEE, Elsevier, Springer, Wiley, De Gruyter and others. His research interests include wireless sensor networks, digital image processing, data mining, smart cities, the Internet of Things, machine learning, big data, digital signal processing, wireless communication, and VLSI.

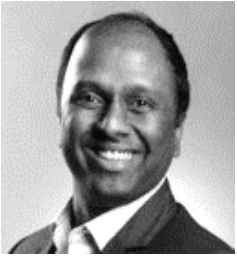

**V. B. Surya Prasath** graduated from the Indian Institute of Technology Madras, India in 2009 with a PhD in Mathematics. He is currently an assistant professor in the Division of Biomedical Informatics at the Cincinnati Children's Hospital Medical Center, and at the Departments of Biomedical Informatics, Electrical Engineering and Computer Science, University of Cincinnati from 2018. He has been a postdoctoral fellow at the Department of Mathematics, University of Coimbra, Portugal, for two years from 2010 to 2011. From 2012 to 2015 he was with the Computational Imaging and VisAnalysis (CIVA) Lab at the University of Missouri, USA as a postdoctoral fellow, and from 2016 to 2017 as an assistant research professor. He had summer fellowships/visits at Kitware Inc. NY, USA, The Fields Institute, Canada, and IPAM, University of California Los Angeles (UCLA), USA. His main research interests include nonlinear PDEs, regularization methods, inverse and ill-posed problems, variational, PDE based image processing, and computer vision with applications in remote sensing, biomedical imaging domains. His current research focuses are in data science, and bioimage informatics with machine learning techniques.